\documentclass[dvips]{acta}
\usepackage{supertabular,lscape,epsfig}
\usepackage{amssymb}
\usepackage{amsmath}
\mathchardef\mhyphen="2D
\usepackage{placeins}%\Floatbarier
\DeclareSymbolFont{ppa}{OT1}{ppl}{m}{it}
\DeclareMathSymbol{\vv}{\mathalpha}{ppa}{'166}

\def\bn{\setbox0=\hbox{0}\hbox{}\hskip\wd0\hbox{}}

\SetPages{1}{1}

\SetVol{69}{2019}

\begin{document}

\begin{Titlepage}

\Title{Variable Stars in M13.  III. The Cepheid Variables and their Relation to 
Evolutionary Changes in Metal-poor BL Her Stars}

\Author{W.~O~s~b~o~r~n$^1$,~~G.~K~o~p~a~c~k~i$^2$,~~H.~A.~S~m~i~t~h$^3$,~~ A.~L~a~y~d~e~n$^4$,~~B.~P~r~i~t~z~l$^5$, C.~K~u~e~h~n$^6$,~~M.~A~n~d~e~r~s~o~n$^3$} 
{$^1$Yerkes Observatory, Dept. Astron. \& Astrophys., U. Chicago, Chicago, IL 60637, USA
%373 West Geneva Street, Williams Bay, WI 53191, USA\\
Dept. of Physics, Central Michigan Univ., Mount Pleasant, MI 48859, USA\\
e-mail:Wayne.Osborn@cmich.edu\\

$^2$Instytut Astronomiczny UWr., Kopernika 11, 51-622 Wroc{\l}aw, Poland\\
e-mail: kopacki@astro.uni.wroc.pl\\
$^3$Dept. of Physics \& Astronomy, Michigan State Univ., East Lansing, MI 48824, USA\\
e-mail: smith@pa.msu.edu\\
$^4$Dept. of Physics \& Astronomy, 
%104 Overman Hall, 
Bowling Green State Univ., Bowling Green, \\OH 43043, USA ~~~~~~~~~~~
e-mail: laydena@bgsu.edu\\
$^5$Dept. of Physics \& Astronomy, Univ. of Wisconsin Oshkosh, Oshkosh WI 54901, USA\\
e-mail: pritzlb@uwosh.edu\\
$^6$Univ. of Northern Colorado, Campus box 126, Greeley, CO 80639, USA\\
e-mail: charles.kuehn@unco.edu}

\Received{Month Day, Year}

\end{Titlepage}

\Abstract{New CCD photometry has been combined with published and unpublished earlier observations to study 
the three Cepheid variables in M13: V1, V2 and V6. The light curve characteristics in $B$, $V$ and $I_{\rm C}$ have been determined and the periods updated. A period change analysis shows all three stars have increasing periods but for V1 and V2 the rate of period increase does not appear to be constant over the 118 years of observation. The observed rates of period increase are in good agreement with the predictions of the Pisa theoretical models with helium abundance $Y = 0.25$. Theory suggests V1 and V6 have masses of $\sim0.57\,M_\odot$ and are in the redward-evolving final stage of the ``blue loop'' evolutionary phase that is produced when helium-shell ignition occurs. The larger period and period change rate for V2 indicate it has a mass of $\sim0.52\,M_\odot$. A study of eighteen metal-poor BL Her stars shows the observed period changes for such objects in general can be reasonably well explained using the predictions from horizontal branch evolutionary tracks. BL Her stars with periods less than $\sim$3 d and relatively large secular period change rates ($dP/dt\approx5-15$ d/Myr) are in the evolutionary stage before He-shell ignition; the remaining cases are stars that have already experienced He-shell ignition.  Moreover, an analysis of crossing time through the instability strip indicates that it is likely that few, if any, BL Her stars have a He abundance as large as $Y = 0.33$.}
{globular clusters: individual (M13) -- stars: individual (M13 Cepheids)}

\section{Introduction}

Observations of pulsating variable stars provide one of the better ways for testing theoretical stellar models and evolution.  Evolutionary effects should produce observable changes in the pulsational parameters of the variables.  Work has generally concentrated on seeking changes in pulsational periods, since periods can be determined with higher accuracy than any other measurable quantity.  The short period Population II Cepheids, or BL Her stars, have been of particular interest in period change studies.  Such stars are believed to be low-mass stars evolving from the horizontal branch toward the 
asymptotic branch as their core helium is exhausted.  Longer period type II Cepheids, including W Vir variables, are usually believed to be asymptotic branch stars undergoing blueward loops due to helium shell flashes as they evolve up the asymptotic branch (Smith \etal 1978, Wallerstein 2002, Catelan and Smith 2015, but see Bono \etal 2016 for a partially divergent view). In either case, the stage of evolution is predicted to produce relatively rapid, and therefore more observable, period changes.

It has recently been proposed that metal-poor short-period Cepheids, such as those found in globular clusters, be called UY Eri variables and BL Her be reserved for near solar-abundance ones (Kovtyukh \etal 2018a), but in this paper we continue to use the BL Her nomenclature for all short-period Cepheids.  The cluster M13 (NGC\,6205 = CL\,1639+365) has three such objects:  V1, V2 and V6.  Early period change investigations of these stars (Osborn 1969, Wehlau and Bohlender 1982) suggested possible small changes for V1 and V6 but a large period increase for V2, the brightest and reddest of the three stars and the one with the longest period.   The V2 result, however, largely depended on two observation sets of dubious quality:  magnitude estimates made visually in 1900 by Barnard (1900a, 1900b) and a very limited number of photographic observations -- only 7 -- by Shapley (1915). More recent work (Smith \etal 2015) confirmed small period changes for V1 and V6 as well as the larger change for V2.  

In this paper, we continue investigation of the M13 Cepheids, including period changes, using recent CCD-based observations combined with previously unpublished photoelectric and photographic data (described in Sec.\ 2), which greatly bolster published photometry on these stars.  Together, the combined observational material enables us to update the light curve ephemerides (see Sec.\ 3), to obtain reliable light curves on the standard $UBVR_{\rm C}I_{\rm C}$ system (see Sec.\ 4) and to perform period-change analyses with data of unparalleled duration and coverage (see Sec.\ 5).  This paper is the third in a series of studies of the variable stars in M13 and makes extensive use of the work of Kopacki, Ko{\l}aczkowski and Pigulski (2003; hereafter KKP03) who provided a thorough history of work on the M13 variables to that time.  The series initial paper (Osborn 2000; Paper I) presented positions and mean $UBVR_{\rm C}I_{\rm C}$ values for known and suspected variables and for stars suitable as local comparison stars for the variables.   The second paper (Osborn \etal 2017; Paper II) was an in-depth study of the cluster's red variables, also making considerable use of the KKP03 work.

\section{The Observational Material}

\subsection{CCD Data}

New CCD photometry of the M13 Cepheids has been obtained using telescopes at the Bowling Green State University (BGSU), Michigan State University (MSU), Macalester College (Macal.) and Wroc{\l}aw University Bia{\l}k\'{o}w Station (Bia\l-14) observatories in the period $2003 - 2014$.  In addition, the $V$ and $I_{\rm C}$ observations in flux units of KKP03 made at Bia{\l}k\'{o}w in 2001 (Bia\l-01) have been re-reduced to obtain magnitudes. 

Most details concerning our CCD observations have been presented in Paper II.  Magnitudes were determined using the Alard and Lupton (1998) image subtraction method (ISM) except for the BGSU data of V6 where magnitudes from the \textsc{daophot} profile-fitting reduction package (Stetson 1987, 1994) gave light curves with somewhat less scatter.  Table 1 summarizes the various CCD data sets, denoted for later reference as sets 1 -- 5.  The final three columns show the number of observations available in $B$, $V$, and $I_{\rm C}$ for the three stars.  The individual observations are given in tables accessible through the on-line Appendix.  

\tabcolsep=3pt
\MakeTable{cclcccc}{9.3cm}{New CCD observations of M13 Cepheids}
{\hline
Set& Year range& Telescope& System& \multicolumn{3}{c}{Number of observations} \\
       &                  &                 &             &  V1 & V2 & V6 \\
\hline
\noalign{\vskip2pt}
1 & 2001             &Bia\l-01 0.6-m reflector & $V,I_{\rm C}$       & 341, 321 &342, 322 &325, 304 \\
2 & 2003${}-{}$2010&MSU  0.6-m reflector    & $B,V,I_{\rm C}$  & 49, 49, 31 & 65, 65, 60 &  -- \\
3 & 2004             &Macal. 0.4-m reflector  & $B,V,I_{\rm C}$  & 15, 15, 13 & 13, 14, 14  &  --   \\
4 & 2006${}-{}$2011&BGSU 0.5-m reflector   & $V,I_{\rm C}$          & 83, 100   & 83, 102  & 83, 99 \\
5 & 2014            &Bia\l-14 0.6-m reflector  & $B,V,I_{\rm C}$ & 192, 237, 233& 196, 241, 242 & 177, 235, 228\\
\hline
}

\subsection{Other Observations}

Reliable period-change determinations rely on data extending back in time as far as possible.  We have gathered archival observations dating to 1899.  This material includes both published data and new magnitudes determined by us.  The quality is mixed, ranging from lower-accuracy magnitudes determined from eye estimates on photographic plates -- and even some early visual estimates -- to more reliable photometry determined photoelectrically (Paper I) and from plates measured by an iris photometer (Arp 1955, Demers 1971, Osborn and Fuenmayor 1977,  Pike and Meston 1977, Russev 1973) or microdensitometer (Russev and Russeva 1979, Russeva and Russev 1983, Welty 1985, Paper I).  

The additional material is more completely described in the Appendix.  Table 2 summarizes the published observations (denoted data sets 11 -- 22) while Table 3 summarizes our new data (data sets 31 -- 39).  Again, the final three columns of the tables give the number of observations available in each passband for the three variables. 

\tabcolsep=3pt
\MakeTable{cclcccc}{12.5cm}{Published archival data for the M13 Cepheids.}
{
\hline	
Set& Year range& Telescope& System& \multicolumn{3}{c}{Number of observations} \\
       &                  &                 &             &  V1 & V2 & V6 \\
\hline
\noalign{\vskip2pt}
11& 1899${}-{}$1911&Yerkes 102-m refractor        & visual &94     &268     &  -- \\   	
12& 1914${}-{}$1915&Mt. Wilson 152-m reflector   &pg, pv&7, 3   & 7, 3    &7, 3 \\
13& 1925${}-{}$1938&Babelsberg 122-cm reflector& pg &88     &91      & 89    \\
14& 1932${}-{}$1934&Dom. Astrophys. Obs. 183-cm&pg &27    &27     &26 \\
15& 1935${}-{}$1941&David Dunlap 188-cm reflector& pg &99    & 99    &97 \\
16& 1952                  &Mt. Wilson 152-m reflector &pg, pv &34, 23    & 39, 11  & 43, 34 \\
17& 1962${}-{}$1971&Moscow AZT-2 70-cm reflector & $B$ & 16 & 19 & 20  \\
18& 1967        &US Naval -- Flagstaff 155-cm refl.      &$U,B,V$ & 5, 11, 12 &  5, 11, 12 & 5, 11, 12 \\
19& 1974${}-{}$1981&Belogradchik 60-cm reflector&$B$ & 47 & 54 & 51           \\
20& 1971                 &Mt. Wilson 152-m reflector               &$B,V$ & 57, 59 & 38, 56  & 47, 54 \\
21& 1971${}-{}$1976&U. Western Ontario 120-cm refl.& blue & 28    & 32    & 26 \\
22& 2001${}-{}$2003& 30- and 20-cm catadioptrics, Spain & $V$ &   --  &  --    &  --\\
\hline
\noalign{\vskip3pt}
\multicolumn{7}{p{12.5cm}}{Notes: 11 -- Barnard (1900a, 1900b), Osborn and Barnard (2016), 12 -- Shapley (1915), 13 -- Kollnig-Schattschneider (1942), 14 and 15 -- Sawyer (1942), 16 -- Arp (1955); individual observations not published and numbers of observations are from plotted light curves, 17 -- Russev (1973),  18 -- Demers (1971); Demers' magnitudes are systematically too bright as found by Pike and Meston (1977) and Osborn \etal (2017), 19 -- Russev and Russeva (1979), Russeva and Russev (1983) with unpublished additional measures provided by Russev,  20 -- Pike and Meston (1977), 21 -- Wehlau and Bohlender (1982), 22 -- Violat Bordonau and Bennasar Andreu (2002, 2004), Violat Bordonau, F., Sanchez Bajo, F., and Bennasar Andreu (2005), Violat Bordonau (2015); individual observations not published but CCD light curves and four times of maxima for V2 were given.}
}

In Table 3, data set 31 has the few photoelectric observations.  The remaining sets are observations from photographic plates. Most plate magnitudes are approximately on the $UBV$ system, especially those of data sets 20 and 35 which are based on plates of good scale measured with, respectively, an iris photometer and a PDS microphotometer and transformed to the $UBV$ system through well-observed non-variable stars outside the crowded regions of the cluster. 

Finally, we made some use of observations of the All-Sky Automated Survey for Supernovae (ASAS-SN, see  Shappee \etal 2014a,  Shappee \etal 2014b, Jayasinghe \etal 2018).  ASAS-SN $V$- and $g$-band observations from 2016 -- 2018 are available on-line ({\it http://asas-sn.osu.edu/\/}).  Entering the coordinates of our variables (KKP03) into the Sky Patrol database search tool, we identified all three stars and downloaded their photometric observations (denoted as data set 40 in Table 3).  We found, however, that the relatively large pixel size used in the ASAS-SN survey when used for stars in our crowded globular cluster field severely compromised the photometry and only limited use could be made of these data.

\MakeTable{cclcccc}{9.0cm}{New photoelectric and photographic observations for M13 Cepheids.}
{
\hline	
Set& Year range& Telescope& System& \multicolumn{3}{c}{Number of observations} \\
       &                  &                 &             &  V1 & V2 & V6 \\
\hline
\noalign{\vskip2pt}
31& 1983 &Lowell 183-cm refl. (photoel.)        & $U, B, V$    &    --         &5, 5, 5  &6, 6, 5   \\
\noalign{\vskip2pt}
32& 1900${}-{}$1920&Yerkes 102-cm refractor        & pv    &  17         &  17   & 16  \\	
33& 1949&Yerkes 102-cm refractor                          & pv     &  19        & 19   & 19     \\
34& 1949 & McDonald 208-cm reflector                  & pg, pv  & 15, 14 &15, 14 & 15, 14  \\
35& 1964${}-{}$1983&USNO 1.55-m reflector         &$U, B, V$ & 1, 89, 26  &1, 91, 26  & 1, 89, 25  \\
36& 1967${}-{}$1968&Yale 102-cm reflector            &$B$     &  19     & 20     & 19 \\
37& 1976${}-{}$1980&MSU 0.6-m reflector             &$B, V$  & 93, 2 & 97, 2    &84, 1      \\
38& 1976${}-{}$1989&Yerkes 102-cm reflector       &$U, B, V$  &4, 82, 10&4, 85, 11 & 1, 85, 10     \\
39& 1988&Central Michigan 36-cm refl.                  &pg          &   --      &  1     &   --   \\
40& 2016${}-{}$2018& ASSAS-SN                     &$g,  V$ & -- & -- & -- \\
\hline
}

%\begin{table}[t]
%\setlength{\tabcolsep}{2.5pt}
%  \caption{New photographic observations for M13 Cepheids.\\
%  \label{tab:landscape}
% \begin{tabular}{cclcccc}
%   \hline
%Set& Year range& Telescope& System& \multicolumn{3}{c}{Number of observations} \\
%       &                  &                 &             &  V1 & V2 & V6 \\
%\hline
%31& 1983 &Lowell 183-cm refl. (photoel.)        & U, B, V    &             &5, 5, 5  &6, 6, 5   \\
%   &                                                                        &         &             &       & \\  
%32& 1900${}-{}$1920&Yerkes 102-cm refractor        & pv    &  17         &  17   & 16  \\	
%33& 1949&Yerkes 102-cm refractor                          & pv     &  19        & 19   & 19     \\
%34& 1949 & McDonald 208-cm reflector                  & pg, pv  & 15, 14 &15, 14 & 15, 14  \\
%35& 1964${}-{}$1984&USNO 1.55-m reflector         &U, B, V & 1, 89, 26  &1, 91, 26  & 1, 89, 25  \\
%36& 1967${}-{}$1968&Yale 102-cm reflector            &B     &  19     & 20     & 19 \\
%37& 1976${}-{}$1980&MSU 0.6-m reflector             &B, V  & 93, 2 & 97, 2    &84, 1      \\
%38& 1976${}-{}$1989&Yerkes 102-cm reflector       &U, B, V  &4, 82, 10&4, 85, 11 & 1, 85, 10     \\
%39& 1988&Central Michigan 36-cm refl.          &pg          &         &  1     &      \\
%\hline
%}
% \end{tabular}
% \end{table} 

The older observations were important to this study in two ways.  First, they were used to supplement the CCD observations which do not fully cover all stars' light curves in $B$.  Second, they allowed us to investigate period changes over an unprecedented span of time. 

\section{Periods 
}

All of the M13 Cepheids have well-determined periods.  We have used the available observational material, which encompasses observations over a century, to improve the periods and investigate period changes.  For each variable, we began doing a period search on the CCD data using the Date Compensated Discrete Fourier Transform method as implemented in the program \textsc{vstar}\footnote{Available at {\it http://www.aavso.org/vstar-overview}.} (Benn 2012) to update the ephemerides from those given by KKP03.   
The new equations are:  
\begin{eqnarray}
T_{\rm max}(V1)& = &2457000.312 + 1.459057~E\\
T_{\rm max}(V2)& = &2457003.837 + 5.111415~E\\
T_{\rm max}(V6)& = &2457001.016 + 2.112890~E
\end{eqnarray}
where $T_{\rm max}$ is the heliocentric Julian date (HJD) of brightness maximum and $E$ is the number of cycles from the more current (2014) reference epoch.  A period search on the recent ASAS-SN data yielded modern periods consistent with those above.

\section{The Light Curves}

\subsection{Photometric zero-point uncertainties}

One goal of this study is to determine reliable light curve parameters --  mean magnitudes, amplitudes and colors -- for the M13 Cepheids on the standard $UBVI_{\rm C}$ system. While good internal precision may now be readily obtained in globular cluster photometry, accurately fixing the zero point to 
place the measures on a standard system is notoriously difficult for many globular cluster variables because of problems caused by crowded fields and varying background from unresolved faint stars. Measures that fail to properly correct for these effects can lead to magnitudes with an offset from the standard system which may be brightness dependent. This is an issue for V1 and V2 which lie at the edge of the dense central region of the cluster and to a lesser extent V6 which has a bright companion. We therefore first attempt to set the zero points of our measures and estimate their uncertainties before discussing the light curves and determining their parameters.  The ASAS-SN observations have been disregarded because the survey's use of a relatively large pixel size in our crowded fields led to light curves with very large observational scatter, distorted shapes and much reduced amplitudes.

Table 4 shows the magnitudes of maximum and minimum light derived from the light curves from our various CCD observation sets as well as those from the two best -- well-calibrated -- sets of photographic photometry: data sets 20 (Pike and Meston 1977) and 35 (see Paper I).   The photographic results  are, of course, less reliable but the light curves for V6 depend on them, especially for $B$, because the CCD data do not fully cover the variation cycle.  Uncertain data are given in italic type. We also show our adopted maxima and minima from intensity-weighted fits to all points in these data sets as discussed below.  In general, the results from the independently determined light curves agree well.  This agreement indicates any zero point errors in our adopted values are less than 0.02 mag.  

\MakeTable{cllcccccc}{12.0cm}{Zero point comparisons for the M13 Cepheids.}
{\hline
Data & Set name&Type &  V1 & V1  & V2 & V2 &  V6   & V6   \\
 set    &  &     & max & min & max &min & max  & min\\
        &  &     & [mag] & [mag] & [mag] &[mag] & [mag]  & [mag]\\
    \hline
\noalign{\vskip2pt}
 & $B$-band                     &             &            &         &          &          &             \\     
5 &Bia{\l}k\'ow &CCD      & \textit{13.8}      &  --  & \textit{12.9} & 14.12 & \textit{14.2}  & --    \\
2 &MSU      &CCD       & 13.70 & 14.98 & 12.93 & 14.10 & -- &--\\
3 &Macal.  &CCD        & 13.74  & 14.96 & 12.90 & 14.16 & -- &--\\
%Russev  &Micr.        & 13.6  & 14.9 & 12.75 & 14.1 &13.95 &  14.99 \\    
35&USNO   &Micr.        & 13.79  & 15.06 & 13.03 & 14.22 &14.18 &  14.98 \\     
20& \underline{Mt. Wilson}   &Iris & \underline{13.82}  & \underline{14.89} & \underline{\textit{13.0}} & 
     \underline{\textit{14.0}} & \underline{\textit{14.1}}     & \underline{14.94}     \\  
%Wehl.     &Iris          & 13.68  & 15.02 & 13.05 & 14.02 &14.05      &14.95     \\  
 & ADOPTED   &               & 13.72  & 15.01 & 12.93 & 14.15 &14.19 &  14.98 \\  
\noalign{\vskip2pt}
 & $V$-band        &              &             &            &         &          &          &             \\     
1,5 & Bia{\l}k\'ow  &CCD       & 13.50 &14.54  & 12.60 & 13.47 & 13.78  & 14.39     \\
4 & BGSU    &CCD       & 13.50 & 14.52 & 12.58 & 13.45  & 13.80  & 14.37 \\
2 & MSU      &CCD       & 13.50 & 14.56 & 12.60 & 13.47  & --   &-- \\
3 & Macal.   &CCD        & 13.52 & 14.56 & 12.63 & 13.52 & --&-- \\
35 &USNO   &Micr.        & 13.52  & \textit{14.46} & 12.74 & 13.50 &13.80 &  14.34 \\     
20 & \underline{Mt. Wilson}   &Iris          & \underline{13.52}  & \underline{14.52} & \underline{12.68} &   
    \underline{13.50}  & \underline{13.84}     & \underline{14.40}     \\  
 & ADOPTED   &               & 13.50  & 14.54 & 12.60 & 13.48 &13.79 &  14.37 \\  
\noalign{\vskip2pt}
 & $I_{\rm C}$-band       &              &             &            &         &          &          &             \\     
1,5 & Bia{\l}k\'ow  &CCD       & 13.17 &13.88  & 12.08 & 12.71 & 13.20  & 13.61     \\
4 & BGSU     &CCD       & 13.16 & 13.90 & 12.05 & 12.68  & 13.20  & 13.62 \\
2 & MSU      &CCD       & 13.19 & 13.93 & 12.07 & 12.73  &  --    &  --  \\
3 & \underline{Macal.}   &CCD       &\underline{13.18} &\underline{13.95} &\underline{12.07} 
    & \underline{12.72} &--      &--     \\
 & ADOPTED   &               & 13.18  & 13.88 & 12.08
    & 12.71 & 13.19 & 13.61 \\  
\hline

}

\subsection{Light curve shapes and parameters }

Figures 1 -- 3 show the $B$, $V$ and $I_{\rm C}$ light curves for the three variables.  The best-quality CCD observations from Bia{\l}k\'{o}w are shown as red crosses, other CCD observations (BGSU, MSU, Macalester) as blue circles, and the photoelectric (data set 31) and well-calibrated photographic observations (data sets 20 and 35) as green squares.  The various observation sets have been rectified to the Bia{\l}k\'{o}w data, that is shifted in phase to account for the period changes and, if necessary, adjusted slightly in magnitude zero point and amplitude to obtain the best fit.   The mean light curves, derived from finite Fourier fits to the combined data, are indicated by solid lines.  For ease in comparing the light curves in the different passbands, for V2 and V6 the $B$ and  $I_{\rm C}$ curves have been shifted closer to the $V$ one by $-0.1$ and 0.3 magnitudes respectively.  

\begin{figure}[t]

\includegraphics{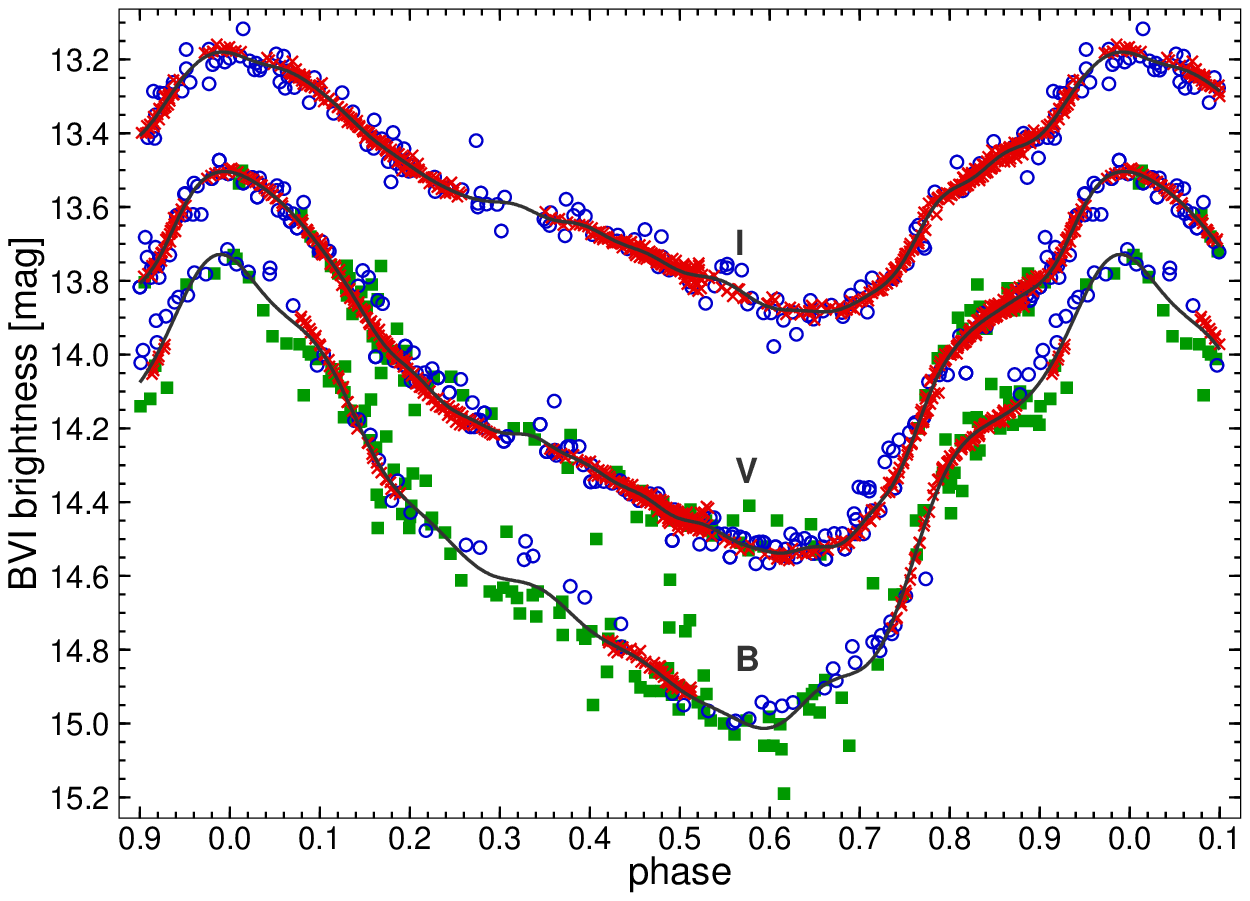}
  \FigCap{Variable 1 light curves for $B$ (bottom), $V$ (middle) and $I_{\rm C}$ (top). Bia{\l}k\'{o}w observations are shown as red crosses, other CCD observations as blue circles, and the photoelectric and well-calibrated photographic observations as green squares.  Solid lines show the finite Fourier series fits to the observations.
}
\end{figure}

\begin{figure}[t]
%\vspace{-4.0cm}
\includegraphics{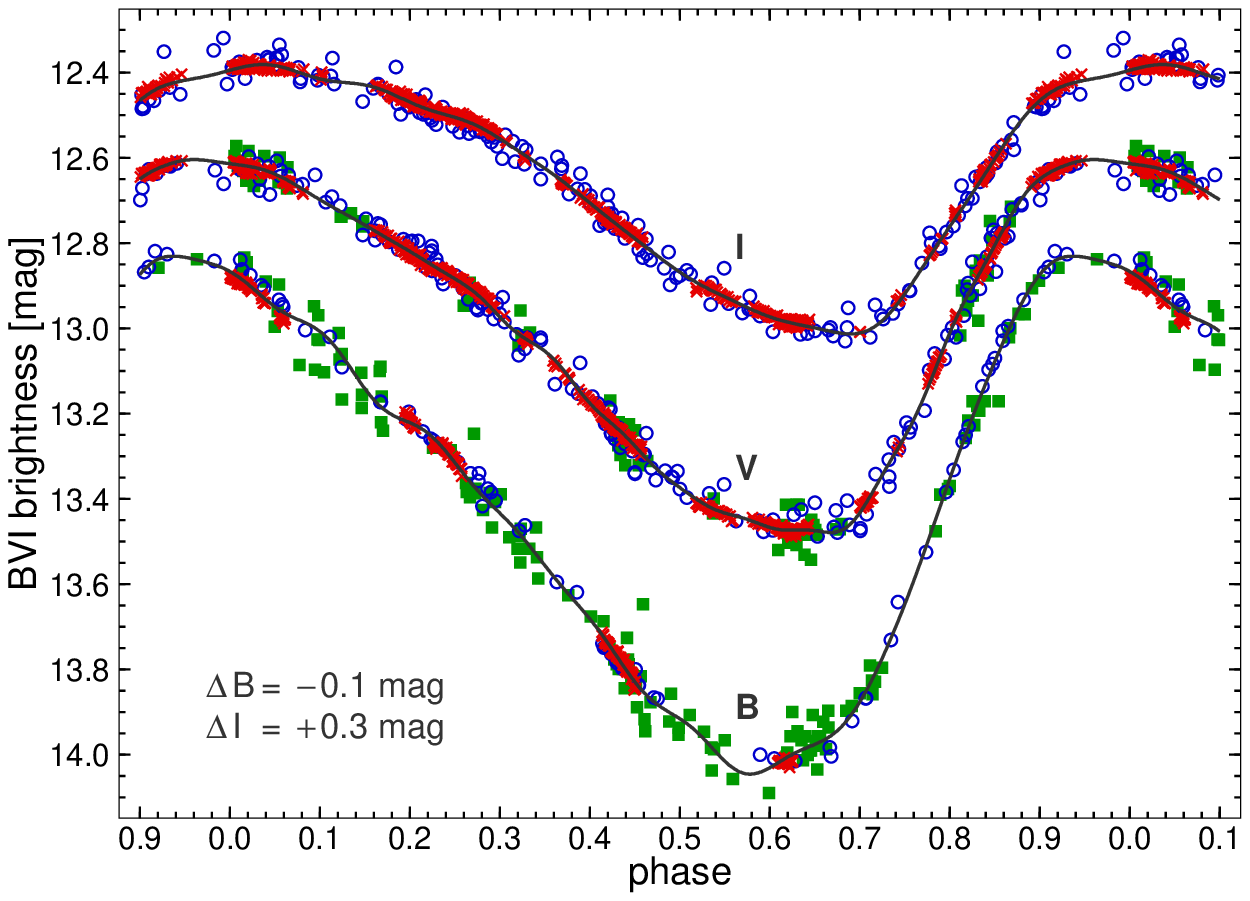}
  \FigCap{Variable 2 light curves for $B$, $V$ and $I_{\rm C}$. Symbols are the same as in Fig.\ 1. For ease in comparison the $B$ and  $I_{\rm C}$ light curves are shifted closer to the $V$ one by $-0.1$ and $+0.3$ mag respectively.}
\end{figure}

\begin{figure}[t]
%\vspace{-4.0cm}
\includegraphics{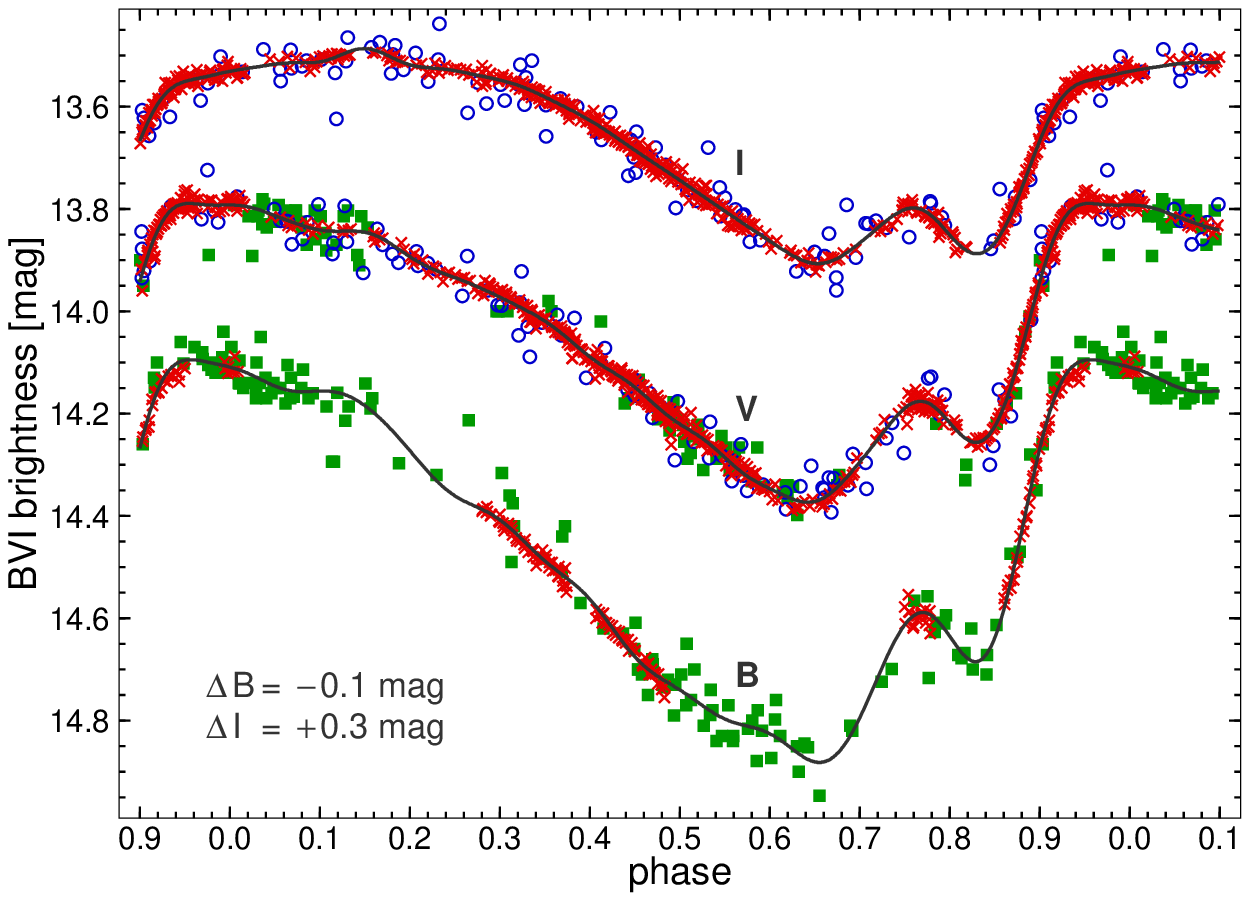}
  \FigCap{Variable 6 light curves for $B$, $V$ and $I_{\rm C}$. Symbols are the same as in Fig.\ 1.   
  For ease in comparison the $B$ and $I_{\rm C}$ light curves are shifted closer to the $V$ 
  one by $-0.1$ and $+0.3$ mag respectively.}
\end{figure}

Various light curve parameters are listed in Table 5.  The table gives for each variable and passband the number of data points used for the Fourier fit, the standard deviation ($\sigma$) of the fit, and the derived maximum and minimum magnitudes, the amplitude of the variation and the magnitude-weighted mean and the intensity-weighted mean magnitudes.  Values for $U$ are included although our observations in this passband are sparse, only from photoelectric and photographic measures and have significant uncertainties.  As discussed above, zero point errors of 0.02 mag or possibly larger between a listed value and the standard $UBVI_{\rm C}$ system may exist.\footnote{As we were finishing this paper's final revision, a preprint of a paper by Deras \etal (2019) became available.  Their independently derived amplitudes and intensity-weighted mean magnitudes in $V$ and $I_{\rm C}$ agree within a few hundredths mag of our results which supports this conclusion.}  V1 and V6 display a distinct bump on the rising branch at phase $\sim$0.80 from maximum.  Modeling such features are important for understanding the physical properties of the stars (see, for example, Keller and Wood 2006 and Bono \etal 2002).

\MakeTable{ccrcccccc}{12.0cm}{Light curve parameters of M13 Cepheids.}
{
\hline
Var& Band & Points & $\sigma$  & Maximum & Minimum &  Amplitude & Avr (mag) & Avr (int)   \\
              &       &              & [mag] & [mag]   & [mag]    & [mag] & [mag] & [mag] \\
    \hline
\noalign{\vskip2pt}
V1& $U$  & 10   & 0.12\bn    & 13.92\bn & 15.42\bn & 1.50\bn & 14.71\bn & 14.67\bn \\
  & $B$  &  438 & 0.037  &13.728 & 15.013  & 1.284 & 14.462 & 14.385  \\
  & $V$  &1004 & 0.020  &13.504 & 14.538  & 1.034 & 14.101 & 14.050  \\
  &$I_{\rm C}$  & 778 & 0.015  &13.180 & 13.884  & 0.70 & 13.566 & 13.543 \\ 
\noalign{\vskip2pt}
V2& $U$  & 16   & 0.18\bn  & 13.24\bn & 14.43\bn & 1.19\bn & 13.88\bn & 13.81\bn \\
  & $B$  &  535 & 0.021  &12.931 & 14.146  & 1.215 & 13.530 & 13.454  \\
  & $V$  &1145 & 0.013  &12.604 & 13.479  & 0.875 & 13.020 & 12.977  \\
  &$I_{\rm C}$  & 988 & 0.012  &12.081 & 12.713  & 0.632 & 12.361 & 12.339  \\
\noalign{\vskip2pt}
V6& $U$  & 13   & 0.09\bn  & 14.35\bn & 15.01\bn & 0.65\bn & 14.69\bn & 14.66\bn \\
  & $B$  &  411 & 0.024  &14.195 & 14.982  & 0.787 & 14.582 & 14.551  \\
  & $V$  & 902 & 0.016  &13.789 & 14.372  & 0.584 & 14.062 & 14.044  \\
  &$I_{\rm C}$  & 761 & 0.011  &13.186 & 13.607  & 0.420 & 13.373 & 13.363  \\
\hline
}

Arp (1955) was the first to note that there are differences between the light curves in different passbands.  He found the times of maximum and minimum in the yellow (his $m_{\rm pv}$ light curve) lag behind those in the blue ($m_{\rm pg}$).  Our light curves confirm this effect.  

For V1 the light curves are essentially in phase from the middle of the rising branch to maximum light, after which the $B$ light curve declines more rapidly than the $V$ curve and the $I_{\rm C}$ curve declines more slowly, with $B$ reaching minimum 0.015 (0.02 d) earlier in phase and $I_{\rm C}$ 0.04 (0.06 d) later in phase than in $V$.  

For V2 the $B$ light curve slightly leads the $V$ curve, with maximum $\sim$0.015 in phase (0.07 d) and minimum $\sim$0.08 (0.40 d) earlier. The $I_{\rm C}$ light curve lags behind $V$, reaching maximum $\sim$0.06 later in phase (0.3 d) and minimum occurring $\sim$0.02 later (0.1 d).   

For V6 our $B$ light curve is based largely on more uncertain photographic data which precludes a reliable comparison with the $V$ band although there is the suggestion that the $B$ light curve features may precede those in $V$.  The $V$ and $I_{\rm C}$ light curves show large differences. The two curves are in phase from the bump up the rising branch, but then $V$ reaches a distinct maximum and then declines while $I_{\rm C}$ continues to rise slowly for 0.2 more in phase (0.42 d) before declining to a minimum that occurs 0.02 in phase (0.04 d) later than in $V$. 

A qualitative explanation can be offered as to why light curve shape might depend on wavelength.  At the effective temperatures of our BL Her stars the $B$ and $V$ bands are near the peak of the Planck function, so flux in those bands scales as $R^{2}T_{\rm eff}{^4}$.  However, the infrared $K$ band is on the Rayleigh-Jeans tail of the Planck function and flux scales as $R^{2}T_{\rm eff}{^{1.6}}$. Thus, the change in radius during a pulsation cycle becomes relatively more important compared to changes in effective temperature when going from blue to infrared wavelengths (Jameson 1986) which could produce the observed light curve differences.

The light curves for the three Cepheids show no obvious sign of amplitude or phase modulation akin to the Blazhko effect often seen in RR Lyrae pulsators (Kovacs 2016).  Nor do our extensive observations show any flares similar to the one reported for V2 by Arp (1955).
\vspace{4pt}
\subsection{Unreddened colors and physical data }

The catalogue of Harris (2010) gives $(V - M_{V}) = 14.33$ mag and $E(B-V) = 0.02$ mag for M13 and these were adopted in Paper II.  More recent work (Denissenkov \etal 2017, Barker and Paust 2018) suggest a larger distance modulus (although this makes the observed periods of the RRc variables less in agreement with theory).  Adopting the Denissenkov \etal values of 14.42 mag and $E(B-V) = 0.025$ mag along with the reddening ratios $E(U-B) = 0.72 \,E(B-V)$ (Hiltner and Johnson 1955\footnote{Somewhat different values for the $E(U-B) / E(B-V)$ ratio have been proposed in various studies (see, for example, Burnstein and McDonald 1975, Fitzpatrick 1999, Schlafly and Finkbeiner 2011, Turner 2012), but the low reddening of M13 means the differences from our adopted ratio of 0.72 have negligible effect.}) and $E(V-I_{\rm C}) = 1.35 \,E(B-V)$ (Bergbusch and Stetson 2009), the intensity-weighted mean magnitudes in Table 5 lead to the unreddened colors and absolute $V$ magnitudes shown in Table 6.  In turn, 
$M_{V}$ and the $(B-V)_{0}$ and $(V-I_{\rm C})_{0}$ colors indicate the luminosities ($\log L/L_\odot$) and effective temperatures ($\pm\simeq$100 K) shown using the bolometric correction and color -- temperature relations of Casagrande \etal (2010) and Casagrande and VandenBerg (2014) which are based on the MARCS model atmospheres (Gustafsson \etal 2008).

\MakeTable{cccccccc}{12.0cm}{Unreddened colors and physical data of M13 Cepheids}
{\hline 
Var & $(U-B)_{0}$ & $(B-V)_{0}$ & $(V-I_{\rm C})_{0}$ & $M_{V}$ & $\log L/L_\odot$ & $T_{\rm eff}$ & $M/M_\odot$    \\ 
 & [mag]& [mag]& [mag]& [mag]&& [K]&\\
\hline 
\noalign{\vskip2pt}
V1 & 0.27 &0.31 & 0.48 &$-$0.37 & 2.1 & 6950 & 0.57\bn    \\
V2 & 0.34 &0.45 & 0.61 &$-$1.44 & 2.5 & 6325 & 0.522    \\
V6 & 0.09 &0.49 & 0.65 &$-$0.38 & 2.1 & 6025 & 0.57\bn    \\
 \hline }

\section{Period Changes}

Period change rates for the stars were determined in the usual manner, that is using an $O-C$ diagram that showed how over time the observed epochs of maximum compared to those predicted by a linear ephemeris. The predicted epoch was initially computed using the appropriate ephemeris from Equations 1 -- 3.  For the final calculations we used a period $P_{\rm mid}$ and reference epoch $T_{\rm mid}$ appropriate for the mid-point of the observations, where $P_{\rm mid}$ and $T_{\rm mid}$ are given in Table 7.  These produced somewhat smaller uncertainties for the earliest computed epochs and reduced the $O-C$ range needed for plotting their diagrams, thereby enhancing the visibility of irregular variations in the period-change.  

$O-C$ {values were determined for each observing season for which we had suitable data using two approaches. First, $O-C$ was computed using the heliocentric Julian Date (HJD) of one observation (or sometimes more) when the star's brightness was seen close to maximum in the light curve. Second, when there were sufficient data to derive an acceptable seasonal light curve we determined $O-C$ from the phase shift of the observed light curve relative to a reference curve --  the 2001 CCD Bia{\l}k\'{o}w $V$ light curve for $B$ and $V$ data sets or the Bia{\l}k\'{o}w $I_{\rm C}$ curve for $I_{\rm C}$-band observations.  The phase-shift determination produced an $O-C$ value, but no epoch, so the HJD from the first approach was modified to give an HJD that yielded the phase-shift $O-C$.  The adopted HJD of maximum and $O-C$ for a season were then weighted averages of the values from the two approaches, giving the phase shift result triple weight.
In a few cases only an $O-C$ from the first approach could be obtained or we simply derived an $O-C$ from a published time of maximum.  We note that the ASAS-SN observations allowed us to obtain four recent $O-C$ values for each star, extending the time coverage to 2018. The derived epochs of maximum for each observing season along with the resulting $O-C$ values are given in the Appendix.   There we also give the estimated errors for these quantities and briefly describe how those were determined.

\begin{figure}[t]
\includegraphics{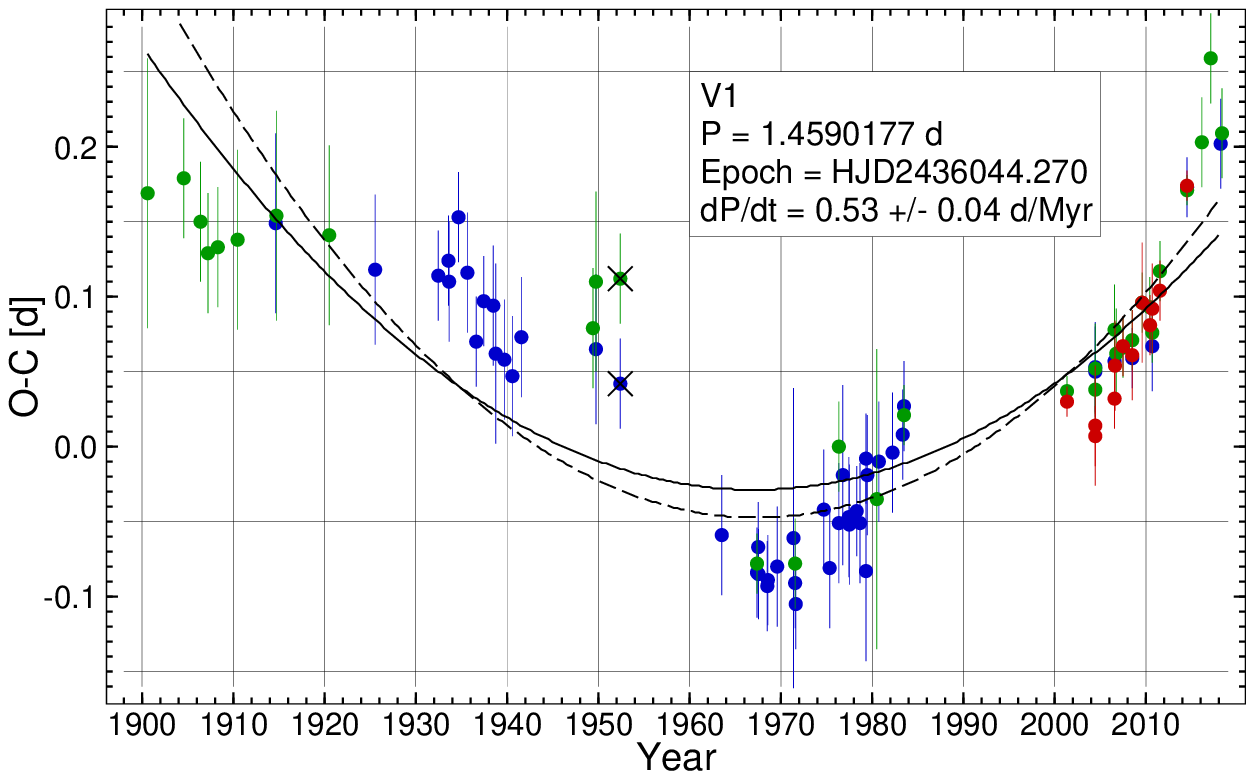}
  \FigCap{$O - C$ diagram for Variable 1.  Green symbols indicate yellow-band ($V$, $pv$, visual) data, blue symbols blue-band ($B$, $pg$, $g$) data and red symbols $I_{\rm C}$-band data. Error bars corresponding to the adopted uncertainties are indicated.  The parabolic fits for unweighted (solid line) and weighted (dashed line) observations are both shown. The two crossed out epochs are those of Arp (1955) which were not used in the fits.}
\end{figure}

\begin{figure}[t]
\includegraphics{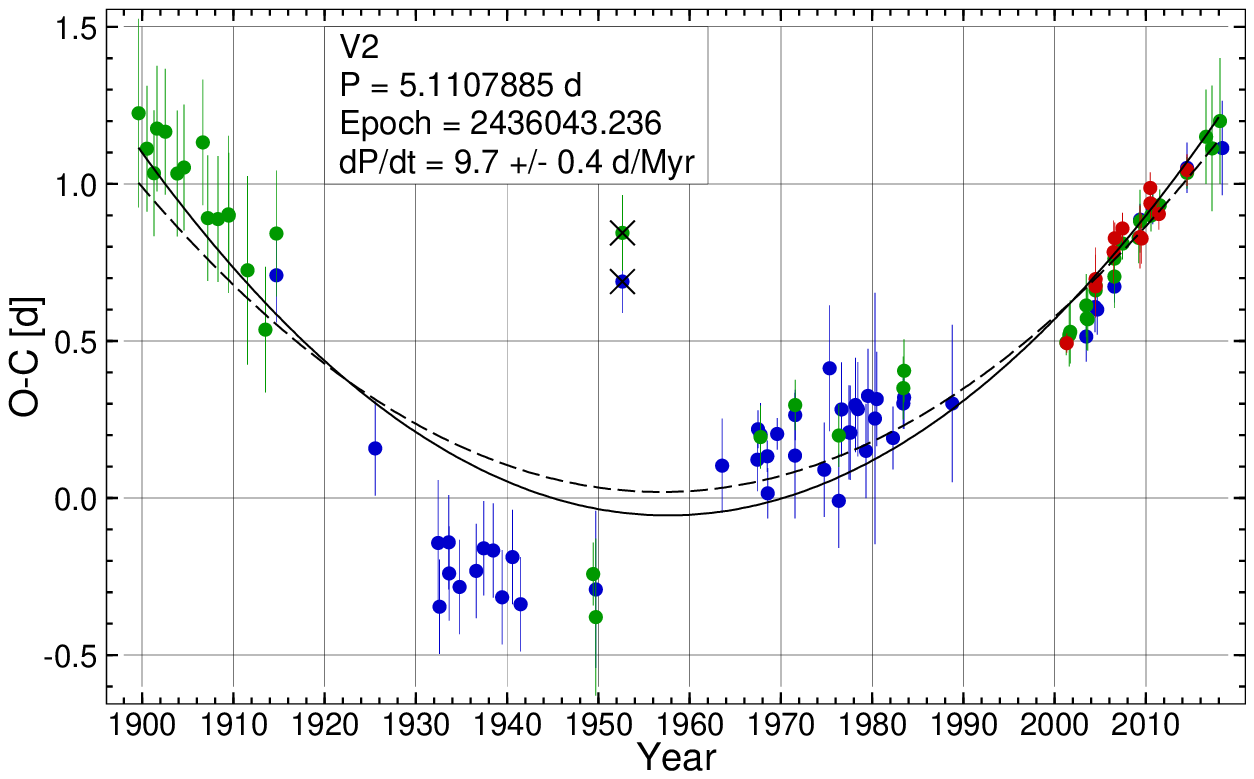}
  \FigCap{$O - C$ diagram for Variable 2.  Symbols are the same as for Fig.\ 4.}
\end{figure}

The $O-C$ versus time diagrams, shown in Figures 4 -- 6, were used to compute the period change rates, 
$dP/dt$.  We used parabolic fits, which assume a constant rate of period change, although the $O-C$ plots for V1 and V2 suggest more irregular changes.  The relevant equations are
%
%\vspace{-6pt}
\begin{equation}
T_{\rm max} = T_{\rm 0} + P_{\rm 0}\, E + a_{3}\,E^{2}~~~~~{\rm and}~~~~~~dP/dt = 2\,a_{3}\,P_{\rm mid}^{-1}
\end{equation}
where  $P{_{\rm 0}}$ is the period at any given reference epoch} $T_{\rm 0}$, $E$ the number of cycles from that epoch and $P_{\rm mid}$ is star's period at the middle of the epoch range of the observations. 
%\textbf{We adopted a reference epoch at the middle of the epoch range, and this $T_{\rm mid}$ along with  $P_{\rm mid}$ and the derived $dP/dt$ values for the three stars are given in Table 7.}  

In the study of RR Lyrae stars and Cepheids, the rate of period change is traditionally given by the quantity $\beta$, which expresses the rate of period change in days per million years (d/Myr) and can be computed from:%
\footnote{ See, for example, LeBorgne \etal (2007).  Our multiplicity factor differs slightly from that given deBorgne et al., apparently because they use sidereal years while we use the standard tropical year.}
%
%\vspace{-6pt} 
\begin{equation}
\beta = 730\times10^{6}\,a_{3}\,P_{\rm mid}^{-1}.
\end{equation}
%\vspace{-6pt}
%
Because of this tradition, the derived rates of period change, $dP/dt$, in this paper are expressed
in units of d/Myr.

We adopted a reference epoch at the middle of the epoch range, and this $T_{\rm mid}$ along with  $P_{\rm mid}$ and the derived $dP/dt$ values for the three stars are given in Table 7.   The epoch data from all passbands were combined, taking care to account for the passband dependency of the light curves.  This was done by shifting the reference epoch for $B$- and $I_{\rm C}$-data by the amounts shown in Table 7 to account for their average light curve shifts relative to the $V$ curve.   We give two $dP/dt$ derived from least squares fits to the $O-C$ values, first assigning them equal weights and then weighting them by 1/error$^2$.  Both fits are shown in each figure. Given the large weight differences between the early and later (CCD) observations, weighting introduces a strong bias toward the most recent epochs. We therefore believe unweighted solutions yield the more reliable determinations of $dP/dt$ and adopt those results for our further analysis.  We note the published epochs of Arp (1955), shown by the crossed out points in Figures 4 --7, gave very discordant $O-C$ values for V2 and V6; we could not find a cause for these discrepancies, so Arp's epochs for all three stars have been disregarded when doing our parabolic fits.

\begin{figure}[ht]
\includegraphics{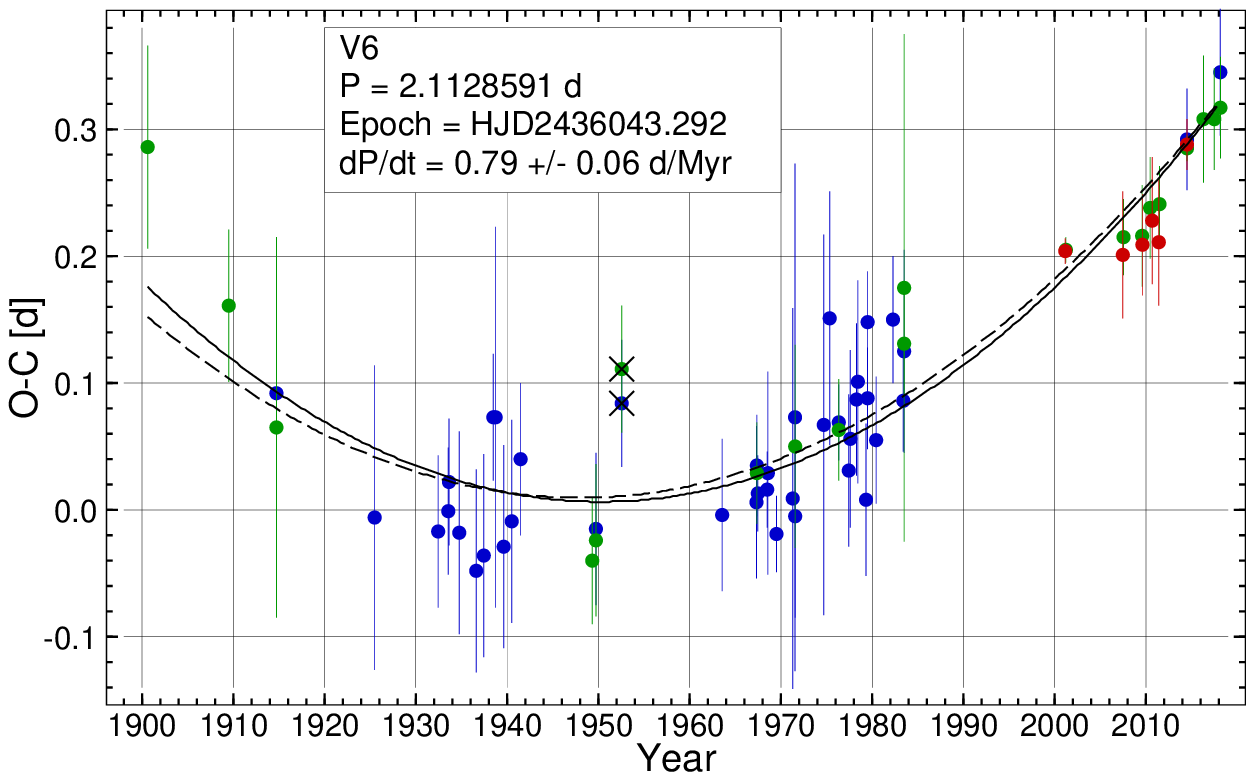}
   \FigCap{$O - C$ diagram for Variable 6. Symbols are the same as for Fig.\ 4.}
\end{figure}

\MakeTable{lccc}{12.0cm}{Derived period change rates of M13 Cepheids.} 
{\hline Parameter & V1 & V2 & V6 \\ 
\hline 
\noalign{\vskip2pt}
$P_{\rm mid}$ [d]& 1.4590177 &5.1107850 & 2.1128591 \\
$T_{\rm mid}$ [HJD]&2436044.270&2436043.236&2436043.292\\
$T_{\rm mid}$ correction for $B$-band [d] &  $+$0.000 & $-$0.070  & $-$0.013 \\
$T_{\rm mid}$ correction for $I_{\rm C}$-band [d] & $+$0.004 & $+$0.280  & $+$0.441 \\
$dP/dt$ (unweighted) [d/Myr]& 0.53 $\pm$ 0.04 & 9.7 $\pm$ 0.4 & 0.79 $\pm$ 0.06 \\ 
$dP/dt$ (weighted) [d/Myr]    & 0.66 $\pm$ 0.04 & 8.4 $\pm$ 0.4 & 0.74 $\pm$ 0.06 

\\ \hline }

All three stars show increasing periods with the rate of increase correlated with the period. The largest change rate is $dP/dt = 9.7$ d/Myr for V2, the brightest variable. The $O-C$ curves for V1 and V2 suggest some rather abrupt period changes rather than a smoothly increasing period. Our results can be compared to those of Wehlau and Bohlender (1982) who found $dP/dt$ values of $0.05\pm0.19$, $18\pm2$ and $0.36\pm0.34$ d/Myr for V1, V2 and V6 respectively.

\section{Comparison to theory}

A number of authors have computed models for metal-poor post-horizontal branch stars and derived the theoretical evolution of BL Her variables found in globular clusters. A recent summary has been given by Neilson, Percy and Smith (2016). Briefly, such stars are believed to be evolving away from the blue horizontal branch towards the asymptotic branch after helium depletion in the core. The models (e.g., Gingold 1976; Bono \etal 1997; Bono \etal 2016, Dotter \etal 2008, 
Dell'Omodarme \etal 2012) predict that blue horizontal branch stars with masses smaller than about $0.51\,M_\odot$ will evolve directly to the white dwarf stage. On the other hand, horizontal branch stars with greater masses become brighter and cooler after core helium depletion, evolving redward across the instability strip at luminosities above that of the horizontal branch (thus producing variables
brighter than a cluster's RR Lyrae stars). This scenario accounts for stars in the portion of the Cepheid instability strip where the BL Her variables are found. Eventually these redward evolving stars leave the instability strip to become red giants on the asymptotic giant branch. Fig.\ 2 in Smolec (2016) illustrates how blue horizontal branch stars evolve to the red through the BL Her portion of the instability strip, based upon the evolutionary models of Dotter \etal (2008) for [Fe/H]${}=-1.0$, $-1.5$, and $-2.0$.  However, as we shall discuss below, more recent calculations indicate the scenario is more complicated,  

\subsection{Theory and the M13 Cepheids}

Fig.\ 7 shows the $(B-V)_{0}$ and $(V-I_{\rm C})_{0}$ color-magnitude diagrams (CMDs) for M13 with the positions of the three Cepheid variables (red symbols) and the cluster's well-observed RR Lyrae stars (green) plotted using their unreddened intensity-weighted mean colors and absolute magnitudes (this paper, Denissenkov \etal 2017).  Superposed are post-horizontal branch evolutionary tracks from the most recent models of the Pisa group (Dell'Omodarme \etal 2012, private communication 2018).  Shown are tracks for masses from 0.52 to $0.58\,M_\odot$ with parameters appropriate to M13: helium abundance $Y = 0.25$, metallicity $Z = 0.0006$ and $\alpha$-elements enhancement [$\alpha$/Fe]${}= 0.3$.   Also indicated are the theoretical blue and red edges of the instability strip for fundamental-mode pulsation (Bono \etal 1997).  Fig.\ 8 shows $(B-V)_{0}$ CMDs with theoretical tracks for $Y = 0.27$ (left panel) and $Y=0.33$ (right panel); tracks with masses up to $0.62\,M_\odot$ are needed to fit the Cepheid positions for $Y = 0.33$.  One sees that increasing $Y$ raises the luminosity of the ZAHB and of the higher mass stars while lowering the luminosity in the instability strip for the least massive stars.

\begin{figure}[t]
\includegraphics{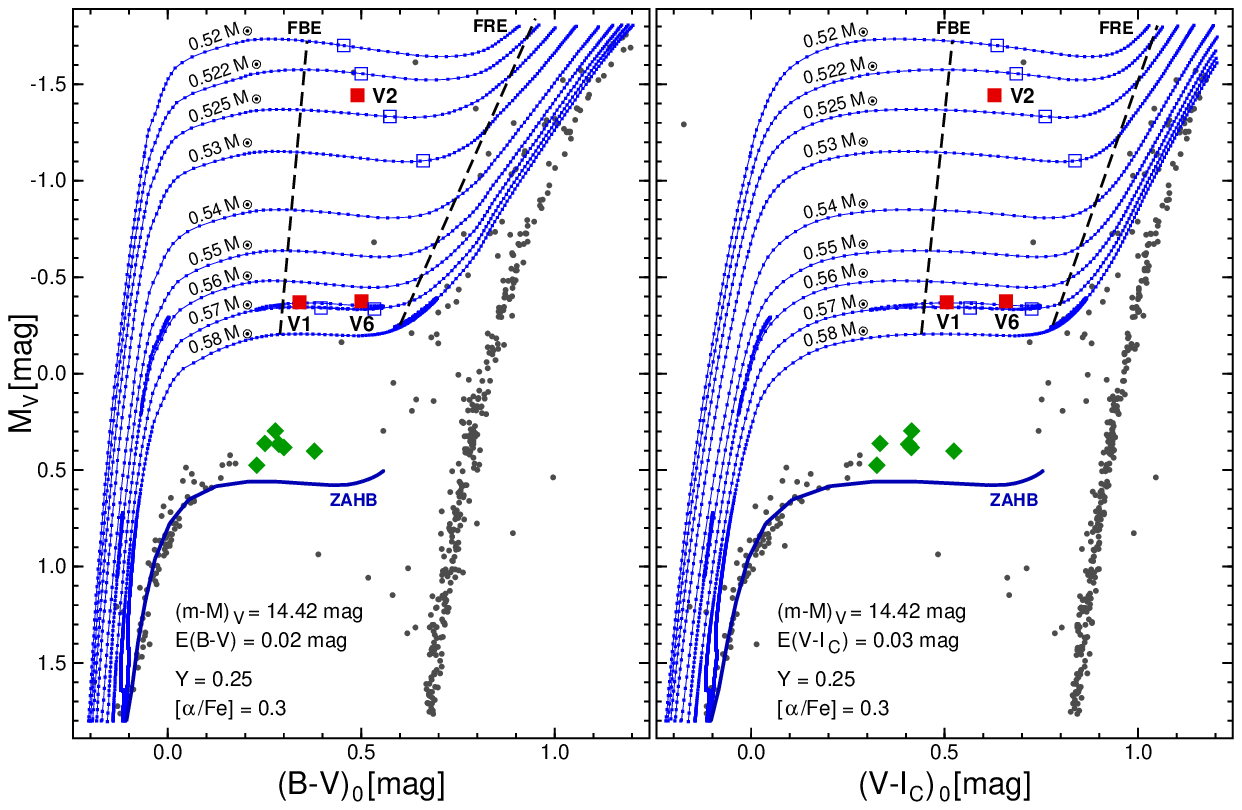}
  \FigCap{Color-magnitude diagrams for M13 using $(B-V)_{0}$ (left panel) and $(V-I_{\rm C})_{0}$ (right panel).  The location of the three 
  Cepheids (red squares) and the best-observed RR Lyraes (green diamonds) are plotted along with several theoretical evolutionary 
  tracks from the Pisa models for $Y = 0.25$.  Blue boxes indicate the theoretical locations for stars with the pulsational periods
  of V1, V2 and V6 and dashed lines indicate the predicted blue and red boundaries of the instability strip (FBE and RBE, respectively).}
\end{figure}

\begin{figure}[t]
\includegraphics{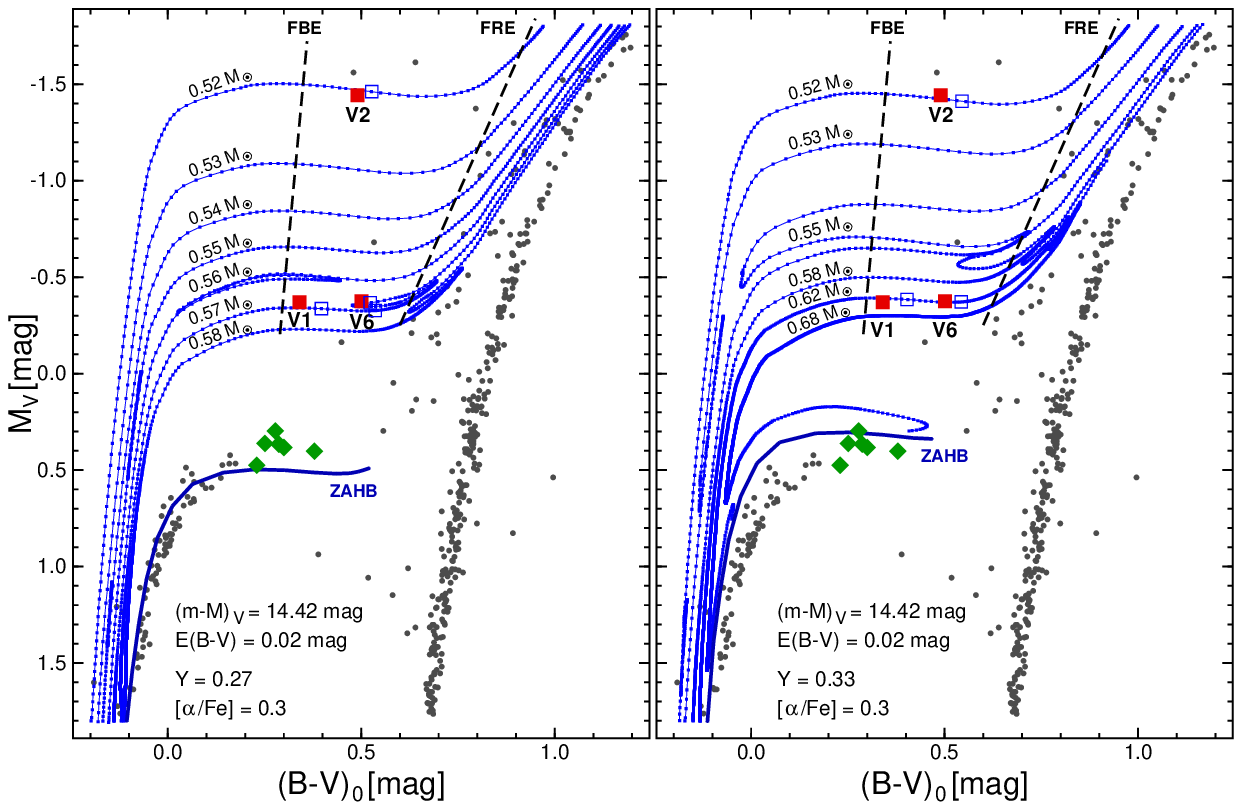}
  \FigCap{$(B-V)_{0}$ color-magnitude diagrams for M13 with Pisa model tracks for Y = 0.27 (left panel) and Y = 0.33 (right panel).  Symbols are the same as in Fig.\ 7.}
\end{figure}

The recent Pisa models indicate that the ignition of shell He burning after the exhaustion of core helium causes loops in the model tracks as the internal structure is reconfigured.  These ``blue loops'' occur at different locations along the tracks depending on $Y$ and $M$, and for certain values can enter the instability strip as seen in Figs.\ 7 and 8.  For a given helium content, the loop occurs further and further along the track, and redder, as stellar mass increases, finally entering the instability strip for the highest masses.  Increasing $Y$ moves the loop farther along the track, amplifies it and increases it in luminosity.
Not shown here, but also used for this analysis, were other CMDs showing the tracks for [$\alpha$/Fe]${} = 0.0$ in place of 0.3.  The change in [$\alpha$/Fe] has little effect on the locations of the tracks.  

For the purposes of the further analysis we separate every evolutionary track into three phases based on the ``blue loop'': the pre-loop first redward evolution (FRE) phase that starts at the ZAHB, the loop-produced blueward evolution (BE) and the second redward evolution (SRE) phase that finishes the loop and terminates at the AGB stage.

The locations of the BL Her stars in Fig.\ 7 and Fig.\ 8 indicate a mass of $\sim0.52\,M_\odot$ for V2 irrespective of helium abundance $Y$.  Masses of 0.57 $M_\odot$ are predicted for V1 and V6, as given in Table 6, but we note their masses could be significantly greater if their helium abundances were to be large compared to what is traditionally thought.  
%Further discussion of the stars’ possible masses and helium abundances is in the following section.
The blue boxes along the upper tracks indicate where a star with the pulsation period of V2 is predicted to lie based on the pulsational relation of Marconi \etal (2004). The predicted pulsation locations corresponding to the observed periods of V1 and V6 are similarly indicated, but two blue boxes are seen near V6 in the $Y = 0.27$ diagram (left panel of Fig.\ 8) -- one if the star is assumed to be on its initial crossing of the instability strip (FRE phase) and a second box (at higher luminosity) if the star is assumed to be evolving redward (SRE) after its blueward loop. There are similar dual FRE and SRE pulsational possibilities for V1 and V6 in the Y = 0.25 CMD, but for both stars the two cases have nearly the same luminosity and they are indistinguishable in Fig.\ 7.   Which of the three evolutionary stages are possible for our variables is discussed in more detail in the following section. For now we note that pulsation theory and observation agree quite well even though for each of the three Cepheids the predicted location is slightly redder than our observed one.

\subsection{Predicted period change rates}

The theoretical evolutionary tracks also permit calculation of the expected period change rate as a star crosses the 
instability strip. Wehlau and Bohlender's theoretical $dP/dt$ values were based on the models of Gingold (1976) and 
Sweigart and Gross (unpublished) which are now outdated.  The presence of loops complicates the period change calculation.  
When there is a loop that enters the instability strip, a star's pulsational period is predicted to first increase as the 
star initially evolves redward into the instability region (FRE phase), then decrease when the star evolves blueward along the loop
(BE phase) and then again increase when the star loops back redward across and out of the instability strip (SRE phase).  

Period-change rates $dP/dt$ were computed for the Pisa models falling into the instability strip. They were defined as $\Delta P/\Delta\tau$, where $\Delta P$ is the difference in calculated periods between a given model and the previous one of the evolutionary track, and $\Delta\tau$  is the corresponding change of the evolutionary age. The boundaries of the instability strip for fundamental pulsations were fixed with Bono et al. (1997) equations 2 and 3 for the blue edge and the strip width, respectively. The theoretical relations for the fundamental period of BL Her stars as functions of mass, luminosity and effective temperature, derived by Marconi et al. (2004, their equation 1a), were used to determine pulsation periods for unstable models. 

The result of the above computations is illustrated in Fig.\ 9 which shows the predicted values of period change rate 
$dP/dt$ as a function of period $P$ for different masses $M$ and helium abundances $Y$.  Models with $Y = 0.25$ are shown as red squares, those with $Y = 0.27$ as green squares and $Y = 0.33$ as blue ones.  The models used [$\alpha$/Fe]${}= +0.3$ and mixing length parameter ML${}= 1.7$ but a change in these parameters has little effect on the diagram.

\begin{figure}[t]
\includegraphics[width=\linewidth]{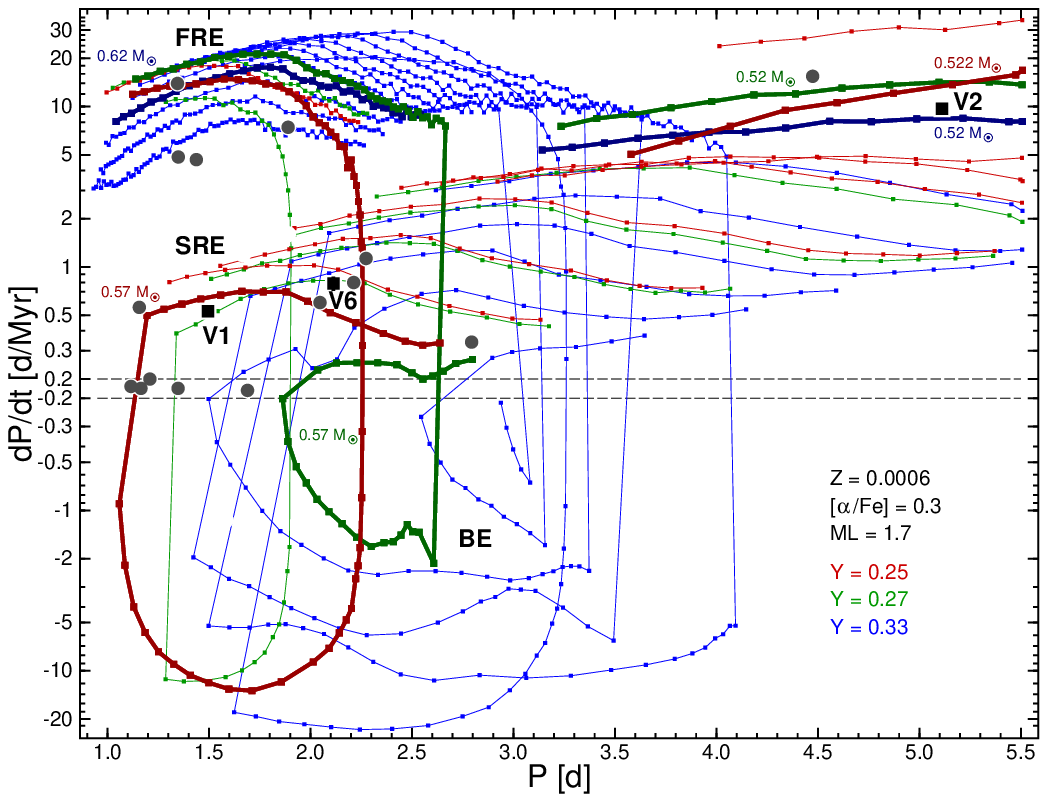}
 \FigCap{Theoretical rates of period change $dP/dt$ as a function of period $P$ calculated from the Pisa stellar evolution models.  
 Models with $Y = 0.25$ are shown as red squares, with $Y = 0.27$ as green squares and with $Y = 0.33$ as blue ones. Models for  the same $M$ and $Y$ are connected by lines of the appropriate color, and the regions corresponding to the three phases of 
 evolution produced by loops are labeled with FRE, BE and SRE (see text). Tracks with masses corresponding to the locations of the M13 Cepheids in the CMDs are emphasized and the mass indicated with the appropriate color. The positions of M13 V1, V2 and V6 (black squares), along with other Type II Cepheids (gray circles), from their observed period change rates are shown.}
\end{figure}

Models for the same $M$ and $Y$ are connected by lines of the appropriate color; the lines can be considered evolutionary tracks in the 
$P$ -- $dP/dt$ plane.  The plotted tracks are for masses from 0.52 up to $0.58\,M_\odot$ for $Y = 0.25$ and 0.27 and up to $0.65\,M_\odot$ 
for $Y = 0.33$.  The tracks with masses that are in agreement with the positions of the M13 Cepheids in the CMDs (i.e., 
$\sim0.52\,M_\odot$ for V2 and 0.57 or $0.62\,M_\odot$ for V1 and V6, see Figures 7 and 8) are emphasized using thick lines and labeled with mass, the label having the same color as the appropriate track. The sections of the diagram corresponding to FRE, BE and SRE stages of evolution are marked. Note that the logarithmic scale of the $dP/dt$ axis exaggerates the differences between the tracks for small $dP/dt$ values. 

\MakeTable{cccccrcrcr}{12.5cm}{Theoretical period change rates of M13 Cepheids.} 
{\hline Var & Period &$dP/dt$ &Evol. &\span\omit\hfill $Y = 0.25$\hfill& \span\omit\hfill $Y = 0.27$\hfill& \span\omit\hfill $Y = 0.33$\hfill \\ 
     & &  &phase &$M/M_\odot$ & $dP/dt$ & $M/M_\odot$ & $dP/dt$ & $M/M_\odot$ & $dP/dt$\\
     & [d]& [d/Myr] & & & [d/Myr] &  & [d/Myr] &  & [d/Myr]\\
\hline 
\noalign{\vskip2pt}
V1 & 1.46 & 0.53 & FRE & 0.57  &   13.9\bn& 0.57 &   19.0\bn& 0.62 & 14.2\bn\\
     &         & & BE  &       &$-$11.2\bn&      &     --\bn\bn&   & --\bn\bn\\
     &         & & SRE &       &    0.6\bn&      &     --\bn\bn&   & --\bn\bn\\
\noalign{\vskip2pt}
V6 & 2.11 & 0.79 & FRE & 0.57  &    6.4\bn& 0.57 &   15.0\bn& 0.62 & 13.3\bn\\
     &         & & BE  &       & $-$7.1\bn&      & $-$1.1\bn&      & --\bn\bn\\
     &         & & SRE &       &    0.6\bn&      &    0.2\bn&      & --\bn\bn\\
\noalign{\vskip2pt}
V2 & 5.11 &  9.7 & SRE & 0.522 &   13.6\bn& 0.52 &   14.2\bn& 0.52 & 8.4\bn\\
     &         & & SRE & 0.525 &    4.6\bn&  -- &    --\bn\bn& -- & --\bn\bn\\ 
\hline }
 
Using the track for $M = 0.57 M_\odot$and $Y = 0.25$ (heavy red line in Fig.\ 9) as an example, one can see the theoretical 
changes of the pulsational period as a star evolves across the instability strip.  The track starts in the upper left portion of 
the diagram with a period of 1.1 d that is increasing at a rate of $dP/dt \approx 12$ d/Myr.  As the star evolves redward across the  strip $P$ increases to 1.6 d and $dP/dt$ to about 15 d/Myr, after which the rate of period increase slows as $P$ grows to 2.25 d.   At this point He shell burning starts to significantly affect the star's internal structure and it begins to evolve back toward higher 
temperatures (the ``blue loop'' effect). The period now decreases, which is reflected by negative $dP/dt$ values.  The fastest period 
decrease occurs at $P$ = 1.7 d with $dP/dt \approx -13$ d/Myr.   When the period has declined to 1.1 d it begins to increase again 
as a SRE occurs along the last part of the loop.  The period now increases very slowly, eventually reaching 2.7 d as the star evolves out of the instability strip.  One can see that $dP/dt$ depends significantly on where a star is in its evolution.  Comparing this $Y = 0.25$ track to the similar one for $Y = 0.27$ (heavy green line in Fig.\ 9) shows that helium abundance also significantly affects  the $P$ -- $dP/dt$ relation.

Table 8 summarizes the predicted period change rates for V1, V2 and V6 from their known periods and the masses indicated 
by Fig.\ 7 and Fig.\ 8 for various $Y$ values and different stages of evolution. Taking into account our observed $dP/dt$ 
values from Table 7, shown again in Table 8 and plotted in Fig.\ 9 (labeled squares), one sees that for V1 the only satisfactory match 
is for $M =  0.57\,M_\odot$ and 
$Y = 0.25$ in the SRE evolutionary phase.  The V6 result is similar: the observed $dP/dt$ is also best matched by SRE phase 
for $M =  0.57\,M_\odot$ and $Y = 0.25$.  For V2, its larger period and $dP/dt$ can be matched with any $Y$ from 0.25 to a 
little less than 0.33 but only with low mass tracks -- $\sim0.52\, M_\odot$ -- again in the SRE phase of evolution (the loop occurring well before a track reaches the IS).  While these results constrain the helium abundances of the three stars to close to the canonical value $Y = 0.25$, they do not provide evidence of different $Y$ among them which would confirm or refute recent findings of variations in helium abundance among stars in some globular clusters (VandenBerg, Denissenkov and Catelan 2016, VandenBerg and Dessenkov 2018, Lardo \etal 2018, Kovtukh \etal 2018b).

\MakeTable{lcccrcl}{12.0cm}{Observed period change rates for short period type II Cepheids (BL Her stars).}
{\hline Var &$P$             & $dP/dt$ & Error & $\Delta T$ & [Fe/H] & Sources  \\
                 & [d]       & [d/Myr]    & [d/Myr]  & [yr]                                   &             &       \\
\hline
\noalign{\vskip2pt}
V716 Oph                  &1.116&\bn0.03&--&76&$-$1.64&Diethelm 1996; Kovtyukh \etal 2018\\
$\omega$ Cen V43   &1.157&\bn0.56&0.14&79&$-$1.53&Jurcsik \etal 2001\\	
BF Ser                       &1.165&0.0&--&60&$-$2.08&Diethelm 1996; Kovtyukh \etal 2018\\
CE Her                      &1.209&0.2&--&65&$-$1.8\bn&Diethelm 1996; Harris 1981\\
$\omega$ Cen V92  &1.345&13.94&0.56&100&$-$1.53&Jurcsik \etal 2001\\
XX Vir                       &1.348&0.0&--&72&$-$1.56&Diethelm 1996; Kovtyukh \etal 2018\\
$\omega$ Cen V60  &1.349&\bn4.85&0.92&100&$-$1.53&Jurcsik \etal 2001\\
M15 V1                    &1.438&\bn4.67&0.23&72& $-$2.37&Wehlau and Bohlender 1982\\
M13 V1                    &1.459&\bn0.52&0.09&114& $-$1.53&This paper\\
M22 V11                   &1.690&\bn0.01&0.19&83&$-$1.70&Wehlau and Bohlender 1982\\
M14 V76                  &1.890&\bn7.43&1.0\bn&48&$-$1.28&Wehlau and Froelich 1994 \\
EK Del                      &2.047&0.6&--&61&$-$1.1\bn&Diethelm 1996; Diethelm 1990\\
M13 V6                    &2.113&\bn0.80&0.06&114& $-$1.53&This paper\\
UY Eri                      &2.213&0.8&--&66&$-$1.73&Diethelm 1996; Kovtyukh \etal 2018\\
$\omega$ Cen V61 &2.274&\bn1.13&0.16&100&$-$1.53&Jurcsik \etal 2001\\
M14 V2                   &2.794&\bn0.34&0.3\bn&48&$-$1.28&Wehlau and Froelich 1994\\
$\omega$ Cen V48        &4.474&15.45&--&79&$-$1.53&Jurcsik \etal 2001\\
M13 V2                    &5.111&9.2&0.7\bn&115& $-$1.53&This paper
\\ \hline 
\noalign{\vskip3pt}
\multicolumn{7}{p{11.5cm}}{Note:  
[Fe/H] given for stars in clusters is the cluster value listed in the web version of the Harris (1996) catalog.}
}

\section{Conclusions}

The improved data for the M13 Cepheids presented here permit a more robust comparison with theory.   The stars' positions in the CMDs and in the $dP/dt$ vs.\ $P$ plot are generally consistent with the new post-horizontal branch  evolutionary tracks of the Pisa group (Dell'Omodarme \etal 2012, private communication 2018). There is also reasonable  
agreement with the predicted pulsation periods of Marconi \etal (2004).  

The agreement with theory suggests a more coherent understanding of these stars and their differences. The pulsation equation tells us that the longer period of V2 compared to V1 and V6 is largely due to its higher luminosity. Evolution theory informs us that the higher luminosity of V2 is a consequence of its lower mass compared to V1 and V6.  Lower mass BL Her stars, evolving from the blue horizontal branch, cross the instability strip at a higher luminosity than their higher mass counterparts.   The significant light curve shape and color ($T_{\rm eff}$) differences between V1 and V6, which have similar $dP/dt$ values but somewhat different periods, likely reflect that V6 is a bit more advanced in its evolution than V1 and/or the structural features that produce the evolutionary blue loops are more important. 

These arguments can be extended to BL Her stars in general.  Table 9 lists short-period Type II Cepheids with [Fe/H]${}< -1.0$ for which period changes have been published, including the three M13 stars.  The columns give the star name, its period $P$, the derived period change rate\footnote{Not all authors determined a period change rate by fitting a parabola to $O-C$ values.  For those that did not do so, we have converted the published change in $P$ to the equivalent change per million years for consistency.} $dP/dt$, 
 its error when given, the time span $\Delta T$ over which the period was followed, the star's [Fe/H] and the source.  The stars are listed in order of increasing period.  The stars' positions are also shown in Fig.\ 9 (gray circles).
 
Two trends are obvious for the eighteen stars. First, all the detected period changes are positive, that is the periods are increasing.  
Second, the period changes fall into three distinct groups: twelve shorter-period stars ($P < 3$ d) with small change rates 
($dP/dt< 1.1$ d/Myr), four shorter-period stars with significantly larger changes ($dP/dt > 4.6$ d/Myr) and two stars of 
larger period ($P > 3$ d) with large period increases.  Following the analysis outlined above, the first and third groups correspond to post-horizontal branch stars in the SRE phase of evolution and the second group is composed of FRE stars. Stars of the third group have significantly smaller masses than those in the other two groups.

As can be seen from Fig.\ 9, there is very good qualitative agreement between the observed secular period changes of BL Her stars and recent theoretical predictions of post-HB evolution.  Moreover, the lack of observed negative rates of period change (at least in the period range from 1 to 5.5 d) indicates that most metal-poor BL Her stars have He abundances close to the canonical value ($Y = 0.25$), with $Y$ values as large as 0.33 excluded.

As a more quantitative approach to comparing theory and observation, we calculated crossing times through the instability strip for evolving stars with different $Y$ values. A CT parameter was defined as the sum of the individual crossing times for a given phase (FRE, BE, and SRE) computed from tracks with masses in the range 0.52 -- 0.62$\ M_\odot$ and having [$\alpha$/Fe]${}=+0.3$.  More specifically, for each $Y$ there are eleven tracks in the specified mass range using a mass step of 0.01$\ M_\odot$, and for each one the crossing time portion for each of the three evolutionary phases was computed and then these were added to obtain CT(FRE), CT(BE) and CT(SRE)  for different $Y$ values.  The results are given in Table 10, which also shows the total crossing times CT(total) and the percentages of time spent in each phase.

\def\bn{\setbox0=\hbox{0}\hbox{}\hskip\wd0\hbox{}}
\MakeTable{c@{\hskip14pt}cc@{\hskip14pt}cc@{\hskip14pt}cc@{\hskip14pt}c}{12.0cm}{Theoretical
crossing times (CT) through instability strip.}
{\hline
   $Y$&   \omit\span\hfil CT$\,$[FRE]\hfil&  \omit\span\hfil
CT$\,$[BE]\hfil&  \omit\span\hfil CT$\,$[SRE]\hfil& CT[tot]\\
   & [Myr]& [\%]& [Myr]& [\%]& [Myr]& [\%]& [Myr]\\
\hline
\noalign{\vskip3pt}
  0.25&     1.6& \bn9& \bn2.3& 13& 14.0& 78& 17.9\\
  0.27&     1.1& \bn5& \bn5.8& 25& 16.5& 70& 23.4\\
  0.33&     3.7& \bn6&   34.3& 52& 28.1& 42& 66.1\\
\noalign{\vskip3pt}
  $N_{\rm obs}$&  & 22& & \bn0& & 78& \\
\hline
}
The percentage of crossing time spent in each evolutionary phase should correlate with the percentages of stars in our sample found in those phases. Our results indicate, as shown in the last line of Table 10, that of the eighteen stars in Table 9 four (22\%) are in the FRE stage and the remainder (78\%) in the SRE stage; none show decreasing periods indicative of BE.  Comparison with the theoretical predictions leads to several conclusions.  First, the observed SRE percentage agrees with expectations from theory for $Y = 0.25$ with reasonable agreement for $Y = 0.27$.  The definite disagreement for $Y = 0.33$ effectively rules out larger than expected helium abundances in our sample, which is also supported by the lack of stars with decreasing periods which should be in the majority if $Y\sim0.33$. On the other hand, the fact that {\it no} decreasing periods are seen in our sample of 18 stars significantly disagrees with the prediction that BE-phase stars should be more common than FRE ones irrespective of $Y$.

We regard the lack of quantitative agreement between theory and observation -- particularly for the BE stars -- as suggestive of unrecognized errors in the models used to calculate the theoretical tracks.  However, the possibility that this results from the small sample size, its somewhat heterogeneous nature, non-evolutionary effects having significantly affected our derived period change rates or from some other factor introduces uncertainty into that conclusion.  A detailed study along these lines with a larger sample of stars would be valuable.

\Acknow

We thank the anonymous referee for valuable comments that led to significant improvement in the paper, particularly in the discussion of the $O-C$ analysis.   HS thanks the U.S.\ National Science Foundation for partial support of this work under grants AST0440061, AST0607249, and AST0707756.  Special thanks are due Emanuele Tognelli for providing the updated HB evolutionary tracks of Pisa group.
We also thank R.\ Russev and D.\ Welty for providing their unpublished observations and acknowledge the contributions of students and colleagues at Bowling Green State University, Central Michigan University, Michigan State University and Yerkes Observatory for their  contributions to deriving the new CCD and photographic observations.  This research has utilized the on-line resources of the American Association of Variable Star Observers, the SAO/NASA Astrophysics Data System the SIMBAD database, operated at CDS, Strasbourg, France, and data from the Two Micron All Sky Survey, which was a joint project of the University of Massachusetts and the Infrared Processing and Analysis Center/California Institute of Technology, funded by the National Aeronautics and Space Administration and the National Science Foundation.

\end{document}